# How to choose which explanation to use with students? Discussing the tensiometer with beginning teachers

Laurence Viennot

Matter and Complex Systems (UMR 7057), University of Paris, Paris, France*

*10 rue Alice Domon et Léonie Duquet 75205 Paris Cedex 13

Corresponding author: Laurence Viennot     laurence.viennot@univ-paris-diderot.fr

**Abstract**

Among the many decisions to be made in their teaching practice, physics teachers must decide how to explain each particular phenomenon to students. This study explores student teachers' (*ST*s) decision-making process when presented with several explanations of a physical phenomenon that might be used for teaching purposes. During individual interviews, seven *ST*s were offered three different explanations of how a tensiometer works, all with an accurate conclusion. They were also supplied with a grid of criteria for critical analysis. Following in-depth critical analysis of the three explanations in close interaction with the interviewer, each *ST* was asked to specify the criteria for their choice of explanation, for themselves and for university students. Even after stressing that they valued consistency, some teachers opted for an explanation for students that they had just described as inconsistent and/or incomplete. More specifically, their comments revealed conflicting ideas about the need for consistency and the various forms of simplicity or incompleteness in a given explanation. The results invite more explicit consideration of the complex roles of simplicity or incompleteness in explanations and the importance of this type of interaction during teacher preparation.

## 1 Introduction

With regard to science learning, Reddish and Hammer (2009) stated that 'Students must be prepared to contend with ambiguities, to make sound judgments about what to accept and what to question, to reconsider past assumptions and adapt to new discoveries. (…). In short, they must learn an adaptive expertise – the ability to respond effectively and productively to new situations and new knowledge as it develops.' These ideas are also valid for teacher education. The expression 'adaptive expertise' implies a certain degree of autonomy. The idea that this ability should be preserved and developed in students also applies to teachers (Hawthorne, 1986; Shulman, 1983). Autonomy is crucial for teachers not only because knowledge develops over time but also because, at any given time, explanations used in teaching necessarily differ to some extent from consensual domain expert accounts. As Ogborn et al. (1996) noted,

'Scientific knowledge, as stabilized in the scientific community, has to be radically transformed before its form is fitted to a given act of teaching' (p. 59). This transformation (*didactic transposition* in the French tradition: Chevallard, 1991) implies some decisions on behalf of teachers as well as of academic authorities. The question of how these decisions are made was originally raised by Shulman (1986) in one of his seminal papers on Pedagogical Content Knowledge: 'Where do teacher explanations come from? How do teachers decide what to teach, how to represent it? (…) How does the successful college student transform his or her expertise in the subject matter into a form that high school students can comprehend? When this novice teacher confronts flawed or muddled textbook chapters or befuddled students, how does he or she employ content expertise to generate new explanations, representations, or clarifications?' (p. 8).

This paper addresses Shulman's questions *Where do teacher explanations come from? How do teachers decide what to teach?* in line with his construct of Pedagogical Content Knowledge (PCK). To that end, the reported study explored how beginning physics teachers choose their explanations of a given content. The study takes account of the fact that, beyond being autonomous, 'responding productively to new situations' (cf. supra) requires one to make 'sound judgments about what to accept and what to question'—in other words, to engage in what is commonly known as 'critical thinking'. In choosing an explanation, physics teachers must critically analyse several candidate explanations. This component of the teacher's decision-making process—the critical analysis of written texts—is central to the present investigation. The research reported here drew on explanations that can be found in the academic literature. The limitations of certain explanations commonly used in physics education have been noted by researchers, who have proposed more relevant explanations for the phenomena in question (e.g. Chauvet, 2006; Härtel, 1985; Chabay and Sherwood, 1996), so extending the range of available explanations. The research reported here acknowledges that teachers may have access to several explanations for a given subject, as illustrated by the numerous examples in Viennot and Décamp (2018, 2020) (see also Table 1). Regardless of the observed frequency of these cases, comparing different explanations for the same phenomenon can help analyse teachers' priorities when making pedagogical decisions.

For present purposes, the chosen topic was the tensiometer—a device used to measure surface tensions of liquids—whose functioning is explained in different ways in the relevant literature. To the author's knowledge, no existing science education study has explored how this topic is taught. *ST*s were asked to choose between several explanations based on a critical analysis of each candidate explanation.

The study addresses two questions: which of the proposed criteria—if any—were decisive in making their choice, and what benefit did they see in this process of critical analysis? These questions were addressed in individual interviews with seven *ST*s.

An account of previous research concerning teachers' use of explanations in physics introduces the two research questions of this study (section 2). The paper then describes the grid of criteria used to critically analyse three explanations of the tensiometer (section 3). Those explanations are elaborated in section 4, followed by a detailed description of the experimental setting in section 5. The main results are reported and discussed in sections 6 and 7, and the paper concludes by noting some implications for teacher education (section 8).

## 2 Previous research, rationale, and research questions

The question of how *ST*s decide which explanation to use in their teaching is informed by several lines of research. Many studies have examined this issue as an aspect of argumentation. As Duschl and Osborne (2002) argued, 'Students need to develop a sense of the criteria for claiming this evidence or argument is better than that evidence or argument. (…) Here is where Toulmin's Argumentation Pattern and Walton's presumptive reasoning categories can provide guidance on the quality of reasoning.' (p. 64). With regard to studies employing Toulmin's Argumentation Pattern (Toulmin, 1958), Duschl and Osborne commented as follows: 'In these studies, emphasis is placed on the identification and use of the structural features of arguments - that is premises, initial conditions, claims, data, warrants, backings and qualifiers - and the process of argument rather than its content.' (p.62).

More recently, Braaten and Windschitl (2011) observed that this angle of attack might usefully be complemented by focusing on models of explanation: '(…), it seems a conversation is needed in the field about how particular scaffolding supports for scientific explanations are useful for supporting the development of students' explanatory ideas *as well as* (their emphasis) argumentation practices. Currently, the curricula, prompts, and scaffolding in these studies direct students to use evidence and reasoning to support assertions resulting in well-articulated statements of justified belief, but not necessarily resulting in a scientific explanation consistent with models of explanation seen in philosophy of science.' (p. 658).

In many of the explanations found in physics texts—at least among those referred to here—the prevailing model of explanation is what researchers in this domain call 'covering law' (Braaten and Windschitl, 2011; Hempel and Oppenheim, 1948). These are cases in which established laws support the mathematical representation of persistent patterns observed in nature, within a specified domain of validity. In terms of the structure of argumentation, such explanations are rarely seen to be defective (Velentzas and Halkia, 2018). Nevertheless, any assessment of the value of such an explanation, either for oneself or for teaching purposes, demands a thorough content analysis. This may identify a flaw such as logical incompleteness, where a relevant variable has been disregarded. In particular, Slisko (2006) and Viennot and Décamp (2018) have shown that any assessment of an explanation's consistency with accepted physics knowledge demands critical vigilance. In many cases, the observed flaws relate to content rather than to the structural features of reasoning; for instance, a law of physics may be denied *de facto*, as when a hot air balloon is modelled as an isobaric situation (Viennot, 2006). Such anomalies highlight the need to investigate teachers' critical analysis of physics explanations in specific content areas.

This is all the more crucial as 'critical thinking' has been universally advocated as a pedagogical priority for at least the last three decades (NRC, 1996; Rocard et al., 2007; Osborne & Dillon, 2008; NGSS, 2013). Beyond general injunctions, however, this process is proving difficult to specify and implement. Indeed, a multitude of available studies on critical thinking are devoted to psycho-cognitive aspects such as reasoning bias (e.g. Nickerson, 1998; Henderson et al., 2015) or to the greater or lesser independence of critical thinking from context (cf. 'the critical debate' : Ennis, 1992 ; McPeck, 1981). But how to analyse academic texts that are supposed to present well established scientific explanations about particular physical contents is a much less well-documented question. The results considered here have been obtained recently on the basis of studies focusing on four specific themes of physics (e.g. radiocarbon dating: Décamp and Viennot, 2015; osmosis: Viennot and Décamp 2016), synthesized in (Viennot and Décamp 2018). These authors found that when asked to analyse flawed written physics explanations, in response to questions such as *Do you find this explanation satisfying? Would you add or change something? ST*s tend to refrain from any criticism. In the most common case, the interviewee was unwilling or unable to critique a contestable text until more knowledge was provided about the topic, even if this was unnecessary from a logical standpoint.  This case, a 'delayed critique', seems to involve feelings of incompetence that hinder activation of critical analysis. At the opposite, in cases of 'expert anaesthesia', individuals with good existing knowledge of the topic at hand proved reluctant to critique a contestable text in that domain, as if they inadvertently completed or corrected the text they were analysing. In both cases, the status of an explanation as a 'teaching ritual'– a very contestable explanation that is very common in teaching practice and undiscussed -  is a factor that may impede critical analysis. These results confirm the need to develop and evaluate tools that can help *ST*s to critically analyse explanations found in their academic environment or online. In this respect, the idea of comparing 'candidate models' has long been considered fruitful for physics students; for example: 'One possibility for helping students understand what aspects of models are important to evaluate would be to engineer situations in which students are working with candidate models from others and having more trouble understanding one model than another' (Schwarz et al., 2009, p. 652, quoted by Sikorski et al., 2017, p. 939; see also Etkina and Planinšič, 2015). However, few studies have explored the situation of a teacher faced with several written explanations of a given subject. The present study seeks to close this gap in the existing literature.

Finally, another research domain that addresses Shulman's questions is PCK, 'as a form of teachers' professional knowledge that is highly topic, person and situation specific' (Van Driel and Berry, 2012, p. 26; see also Abell, 2008; Aydin, 2014). Recent developments in this strand of research include models such as the PCK Summit Consensus Model, which detail several components of teacher PCK, including *knowledge of students' understanding of science, knowledge of assessment of scientific literacy, knowledge of instructional strategies,* and *knowledge of curricula* (Gess-Newsome, 2015, p. 31). Similarly, the CoRe tool (Hume and Berry, 2011; Loughran et al., 2006) was proposed as a means of helping teachers to improve their PCK by explicating the 'big ideas' underpinning teaching of a particular topic to particular students in a particular context. Designed to support explicitly content-related teaching decisions, the tool is intended to enhance teacher preparation (Nilsson and Carlson, 2019), a goal shared by the present study.

Other theoretical initiatives within this research tradition seek to characterise PCK at topic level (topic-specific PCK/TSPCK) (Mavhunga, 2020) as distinct from discipline level. To that end, TSPCK specifies five content-specific constituent components of within-topic PCK: *learner prior knowledge, curricular saliency, what is difficult to understand, representations and conceptual teaching strategies* (Mavhunga and Rollnick, 2013). In the terms of this framework, the investigation reported here examines '*conceptual teaching strategies*', with particular reference to '*what is difficult to understand*'.

In contrast to previous studies, the present study considers the possibility that a candidate explanation may be contestable in terms of accepted physics, so extending the range of the teacher's responsibility.

To that end, a short list of quality criteria is introduced as a tool to facilitate both the teacher's assessment of a given explanation and the researcher's exploration of the teacher's decision-making process (see next section). This practical tool can be used to guide the initial selection of explanatory content, without excluding subsequent revisions based on the various components of PCK. In particular, the present study seeks to explore how the selected criteria impact teachers' pedagogical decisions and asks whether a lack of conformity with accepted physics suffices to proscribe an explanation's use in the classroom. To the author's knowledge, no existing study has addressed this question.

The first phase of the interviews involved a critical analysis of three explanations to collect data on *ST*s' decision-making criteria rather than to assess their critical capability, both for practical reasons (interview duration) and to avoid methodological issues of interference between these two factors. The interviewer led the critical analysis with a view to addressing the following question: once a critical analysis has been co-constructed, and all participants arrive at roughly the same point in this regard, which criteria do *ST*s consider decisive for their choice, and why?

As this study provides a *de facto* framework for assessing the advantages and disadvantages of physics explanations, it could also be a useful tool for teacher preparation and teaching practice. The anticipated benefits align with the research questions addressed here.
- *RQ1: When several explanations of a given physical phenomenon are available and have been critically analysed based on a list of criteria, which of these criteria are decisive in STs' decision making when they choose an explanation in their teaching?*
- *RQ2: After making a choice based on multi-criteria analysis of several explanations, how did STs assess the value of this approach for their teaching practice?*

As the participating *ST*s were interacting closely with an interviewer during the first phase of the discussion, it is important to note that these research questions are only a preliminary step towards understanding what *ST*s may do if left alone with an explanation and a list of criteria. It should also be noted that this survey focuses on the first research question; the second provides only preliminary indications.

## 3 A grid for critical analysis of explanatory texts

Physics is a very structured science aiming at a coherent and parsimonious description of the world, a few laws accounting for a large set of phenomena in a specified range of validity (Ogborn, 1997; Jenkins, 2007 ; Thagard, 2008 ; Kitcher, 1981). In this epistemological perspective, some criteria concerning the choice of an explanation, here of a 'covering law' type, seem *a priori* essential. The explanation should arrive at an accurate conclusion, that is be consistent with observed regularities, it should be internally consistent and should not contradict, directly or indirectly, well established laws, the logical development should be valid, finally the explanation should be generalizable to a large domain of validity. These – so to speak 'minimal' – criteria constitute the first items of the list proposed to the participants in this study (Table 1). That said, it was observed in previous studies (Viennot, 2006; Viennot & Décamp, 2020), that, despite their flaws regarding these criteria, certain explanations were favoured almost ritualistically in current teaching practice. The question then arises whether it is because teachers are not aware of these defects, or because other criteria prevail. Some incidental comments collected in previous studies (Viennot & Décamp, 2018) led us to hypothesise that these *ST*s saw some benefit in using certain flawed explanations despite realising their inconsistency. In the absence of any known previous research documenting precisely this type of situation (i.e. choosing to advance an explanation despite its known inconsistency), it was hypothesized that simplicity or mnemonic value might play a role in such decisions, and these were added to the list of criteria employed. This 'short list' (Table 1) helped to ensure that interviews were completed within a reasonable time (i.e. less than 75 minutes). The list can also be seen as a practical tool to assist teachers' preliminary selection of explanations, to be further specified in light of the different components of their TSPCK. In this regard, the fact that the 'short list' of criteria may not refer to some of these components does not mean that they do not inform the teacher's pedagogical decisions. The issue is purely methodological; as only a limited number of aspects could be addressed, the research design could not accommodate all of them. In particular, it is acknowledged that the three explanations considered here differed little in terms of important components such as *learner previous knowledge* or *curricular saliency*. That said, the number of criteria is clearly a limitation of this study.

With respect to simplicity and mnemonic value, and unlike the first six criteria, it is difficult to assess the intrinsic value of an explanation without reference to the person being explained. For this reason, it is explicitly recognized here that, in terms of simplicity and mnemonic value, what is asked of interviewees is their personal judgement.

Table 1. Criteria for critical analysis of explanatory texts during the interviews: the 'short list'

|   | Criterion | Example* |
|---|---|---|
| 1 | Accuracy of the text's conclusion? (which may undermine judgment) | A case of accurate conclusion despite inappropriate modelling: The 'isobaric hot air balloon': with this incoherent hypothesis and a Newtonian balance between Archimedes' up-thrust, weight of the solid parts and weight of the internal air, a correct relationship between internal and external temperatures is easily obtained. |
| 2 | Internal contradiction? | Saying (in the same text) that the pressure below the meniscus in a capillary tube partly filled with liquid is at once greater and smaller than the pressure above the meniscus. |
| 3 | Direct contradiction of a law? | Saying that Newton's third law holds only for equilibrium situations. |
| 4 | Indirect contradiction of a law? | Speaking of a cyclist who accelerates without any friction between him or the cycle and the ground (nor any external force acting on him). |
| 5 | Logical completeness? (that is, explanations in which at least one link that is necessary for a satisfactory explanation is missing) | Arguing that, in radiocarbon dating, the ($^{14}C/^{12}C$) composition of the atmosphere is stable over time despite the radioactive decay of $^{14}C$ *without explaining why*. |
| 6 | Generalisable? | Over generalisation: 'Objects made of a material denser than water don't float on water'. What about steel boats? |
| 7 | Simplicity? | As perceived by the interviewee |
| 8 | Mnemonic value? | As perceived by the interviewee |

*These examples can be found in (Viennot & Décamp, 2020).

It should be noted that these criteria may seem inappropriate for clear-cut decision-making. For instance, it is questionable whether a given explanation can be declared accurate when its result is consensual but does not explicitly refer to precise conditions of application. The criteria chosen may also appear to overlap, for example, failure to take into account a relevant variable (such as the shape of a body in relation to its buoyancy) leads to both logical incompleteness (to explain that objects denser than water cannot float) and a lack of generality (not all steel objects sink; for example, steel boats float). Moreover, incompleteness can take two forms. First, an explanation can be logically incomplete—that is, an essential link of the argumentation may be missing, as in the radiocarbon dating example. However, an explanation may be characterized as incomplete because its prerequisites are not acknowledged. Here the emphasis is on logical incompleteness. More generally, these criteria are intended to guide critical reflection and not to provide a decision-making algorithm.

## 3 The tensiometer: three explanations

To address the research questions, the tensiometer was identified as a useful target problem because three different interpretations in existing textbooks or articles (e.g. Marchand et al., 2011; de Gennes et al., 2005; Berthier & Brakke, 2012) lead to the same commonly accepted result. This eliminated any possibility that a *ST* might discard an explanation because the conclusion was clearly incorrect, without considering the relevance of other criteria. It should be noted that certain explanations found in the academic literature are better (according to one or another criterion) than the three selected explanations (e.g. Hecht, 1996). The explanations used in this study were selected for their relevance to the criteria listed in Table 1. Note that although the selected explanations are taken from academic sources, they are not verbatim quotes. Rather, the key elements are highlighted to define the *types of contents* that *ST*s were asked to choose between, using the term *explanation* for simplicity. Consistently, *ST*s might view some calculation details as inessential once they accepted that appropriate further details were available to them if needed. The explanations refer to the relationship between the force needed to maintain in equilibrium ($F_{equi}$) an object partially immersed in a liquid and the surface tension coefficient of that liquid ($\gamma_{LG}$).

As the basis for the first explanation (*Expl. 1,* Fig. 1), it is useful to understand that the coefficient of interfacial tension $\gamma_{LG}$ is the free energy per unit surface linked to the presence of molecules on the interface — in this case, between liquid *L* and gas *G*. Pulling the cylindrical object vertically by $dh$ results in an area $l.dh$ passing from wet to dry, where $l$ is the length of the solid-liquid-gas contact line (or 'triple contact line'), and the corresponding change of interfacial free energy $dU_{free}$ is $dU_{free} = (\gamma_{SG} - \gamma_{SL}) l\, dh$. According to the so called principle of virtual work, in a quasi-static vertical displacement $dh$, the work done by the force $F_{equi}$ (needed to balance the capillary forces) is such that $F_{equi}\, dh = dU_{free}$. Here $dU_{free} = (\gamma_{SG} - \gamma_{SL}) l\, dh$, therefore $F_{equi} = (\gamma_{SG} - \gamma_{SL}) l$. Alternatively, based on Young's relationship ($\gamma_{SG} - \gamma_{SL} = \gamma_{LG} \cos\theta$, where $\theta$ is the contact angle), it follows that $F_{equi} = (\gamma_{LG} \cos\theta)\, l$. Note that with this explanation, it is not necessary to analyse the internal forces of the cylinder-liquid system. On the other hand, it can be criticized because, being limited to the force needed to balance the capillary forces, it does not mention the role of the weight of the solid or the Archimedes' up-thrust on this solid. A priori, a *ST* has more than enough expertise to formulate such a critique.

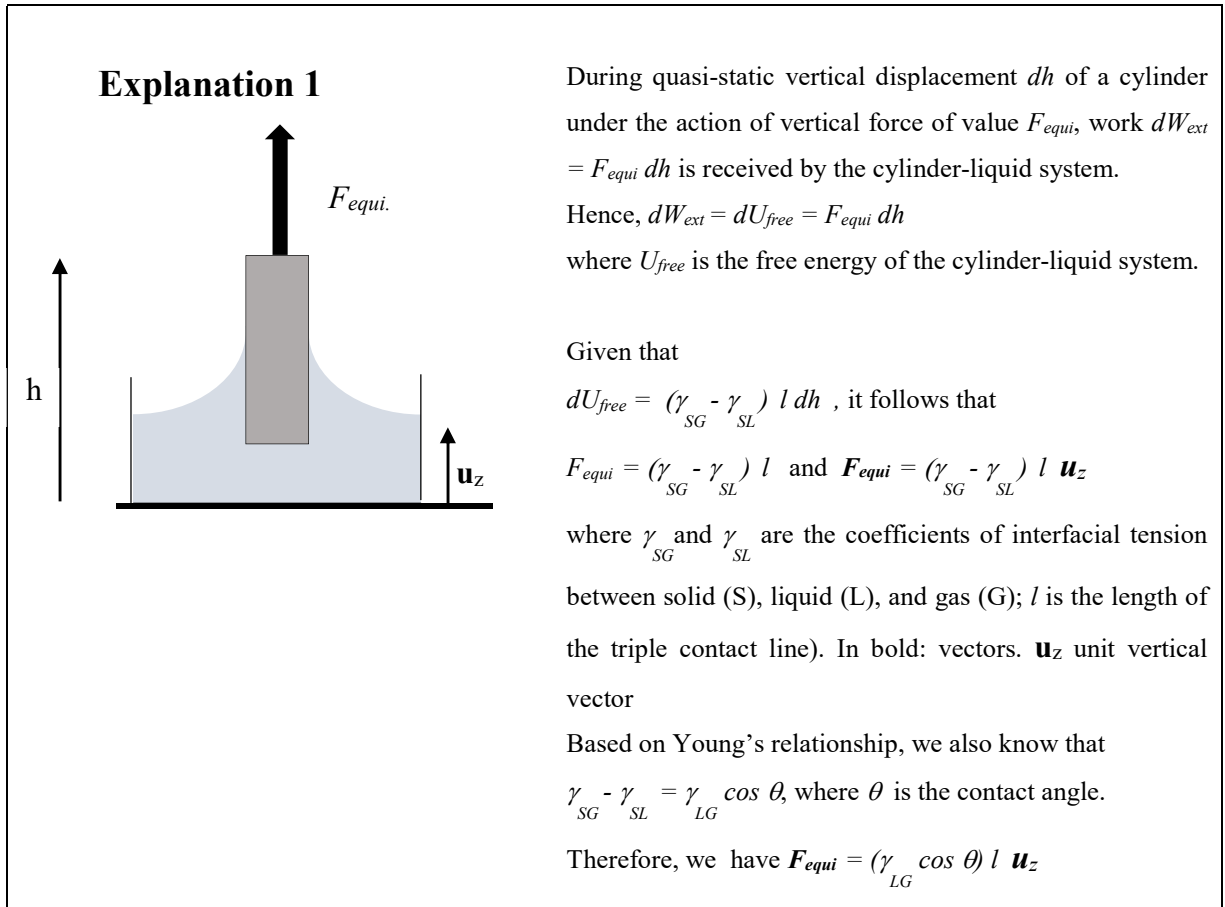

Fig. 1. Main elements of Explanation 1 (*Expl. 1*) of the tensiometer (after Marchand et al. 2011, p. 1000).

Two other explanations are based on an analysis of the forces exerted by the liquid on the solid, which are internal, this time, to the solid-liquid system.

In Explanation 2 (*Expl. 2*, Fig. 2), the starting point is the force by unit length of triple contact line exerted on the solid at the level of this contact line. Its value is $\gamma_{LG}$ and its direction is tangential to the liquid-gas interface at the contact line (downward). A free body diagram indicates the relationship $F_{equi} = (\gamma_{LG} \cos\theta)\, l$.

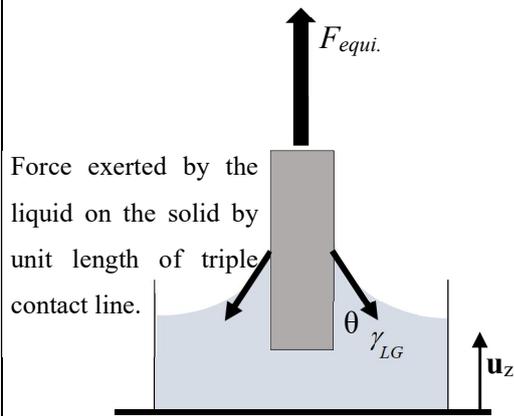

**Explanation 2**

Force exerted by the liquid on the solid by unit length of triple contact line.

The force $F_{equi.}$ needed to hold an object immersed in a liquid at equilibrium can be deduced from the force exerted by the liquid on the solid (by unit length of contact line) of modulus $\gamma_{LG}$ and tangent to the liquid surface at the triple contact line between solid, liquid and gas (angle of contact $\theta$). A free body diagram of the solid indicates the relationship:

$$\boldsymbol{F}_{equi} = (\gamma_{LG} \cos \theta) \, l \, \boldsymbol{u}_z$$

Notations: see Fig. 1.

Fig. 2. Main elements of Explanation 2 (*Expl. 2*) of the tensiometer (after de Gennes et al., p.39; Marchand et al., 2011, p. 1000: Berthier & Brakke 2012, p. 43).

As in the case of Explanation 1, this account does not mention the role of the weight of the solid or Archimedes up-thrust on the solid. On the other hand, the downward forces on the diagram suggest that the local action of the liquid on the solid near the point of contact between the liquid, gas and solid is tangential to the free surface of the liquid. This suggestion of a localised capillary force tangent to the liquid-gas interface is not justified, as if this were self-explanatory. In fact, since intermolecular forces are central forces, this cannot be the case. As shown in Figure 3, the sum of the forces exerted locally by the molecules of the liquid on a given molecule of the solid is directed towards the interior of the liquid.

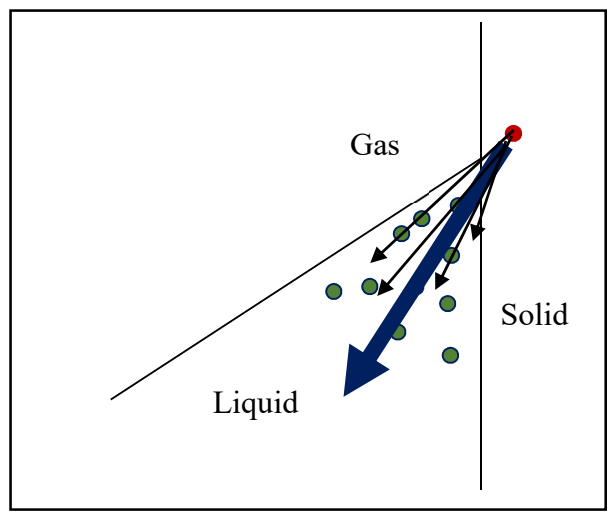

Fig. 3. The forces exerted by the molecules of the liquid on a molecule of the solid near the upper edge of the liquid are central (thin arrows), so the total force exerted locally by the liquid on a molecule of solid (thick arrow) is directed towards the inside of the liquid (after Marchand et al., 2011, p. 1006).

A priori, based on Figure 3, this argument is accessible to any *ST*. However, it contradicts a common teaching ritual, according to which a liquid-solid interface would pull on the solid like a membrane (as in Figure 2).

In Explanation 3 (*Expl. 3,* Fig. 4), the starting point is the value of all forces (by unit length of triple contact line) exerted by the liquid on the solid and distributed on the whole liquid-solid interface. The value of the sum of all these forces is $\gamma_{LG}$, and its direction is the tangent to the liquid-gas interface at the contact point between solid, liquid, and gas (downward), as in Figure 2. Regarding the forces exerted by the liquid on the cylinder in this explanation, their directions are justified based on direct intermolecular attraction near the upper edge of the liquid and curvature of the solid-liquid interface near the bottom of the cylinder. Their values are explained to *ST*s based on (Marchand et al., 2011, p. 1006). As in *Expl. 2*, a free body diagram (Fig. 4) of the solid indicates the relationship: **$F_{equi}$** = $(\gamma_{LG} \cos \theta) \, l \, \boldsymbol{u}_z$ . But, contrary to *Expl. 2*, there is no suggestion that the total force exerted by the liquid on the solid is localized on the triple contact line. Like *Expl. 2* and *Expl. 1,* this explanation does not mention the role of the weight of the solid or Archimedes' up-thrust on the solid.

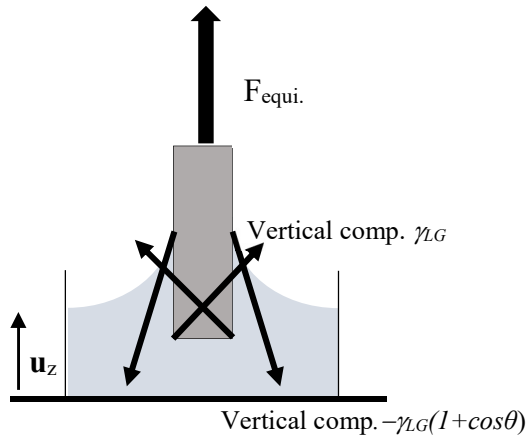

| Explanation 3 | The force $F_{equi}$ needed to hold an object immersed in a liquid at equilibrium can be deduced from the forces exerted by the liquid on the solid (by unit length of triple contact line) at the liquid-solid interface. Their directions can be justified (direct attraction near the gas-liquid wedge, curvature of the solid-liquid interface near the bottom of the cylinder). |

For a cylinder, the diagram indicates the values of the vertical force components exerted by the liquid on the solid (by unit length of triple contact line) corresponding respectively to

- a local attraction near the liquid corner . $-\gamma_{LG}(1+cos\theta)$ and
- the curvature of the solid-liquid interface $\gamma_{LG}$

Hence a free body diagram on the solid indicates

$$\mathbf{F}_{equi} = (\gamma_{LG} \cos \theta) \, l \, \mathbf{u}_z$$

Notations: see Fig. 1

Fig. 4. Main elements of Explanation 3 (*Expl. 3*) of the tensiometer. Justifications for the values of the capillary forces (Marchand et al. 2011, p. 1006) are provided to the *ST*s upon request.

All three explanations lead to the same conclusion, which is accurate regarding the capillary forces in play, disregarding Archimedes' up-thrust on the cylinder and its weight, which are not accounted for in any of these explanations. In all three cases, no special expertise in the field would be needed to formulate a critique of the explanation's incompleteness (e.g. regarding the role of the weight of the solid and Archimedes up-thrust), or of the unjustified and irrelevant location of a force. In short, certain observed limitations in *ST*s' critical analyses cannot be directly attributed to a lack of background conceptual knowledge.

## 4 The experimental setting

*The interviewees' background*

All of the interviewees were close to the end of the final year of a master programme for secondary-level teachers of physics and chemistry. During the programme, *ST*s were required to teach part-time. All participants had attended courses that included capillary forces, surface tension and interfacial coefficients, angles of contact of liquids with a solid and Young's relationship. As might be expected, their recall was sometimes imprecise, but any complementary information they requested was provided by the interviewer at any point throughout the interview.

*The interviews*

The mean interview duration was 72 minutes. Following an introduction outlining the goal of the interview, the participant were presented with the three explanations in question under written form as in Figures 1, 2 and 4. They were informed that all three explanations lead to the same conclusion—that is, the established relationship between the force to be exerted on a suspended object partially immersed into a liquid to ensure equilibrium ($\boldsymbol{F}_{equi}$) and the surface tension coefficient of this liquid ($\gamma_{LG}$). Any relevant information was provided as necessary—for instance concerning Young's relationship. The interviewer stressed that the issue of interest was the interviewee's opinion about these explanations and not their knowledge of the domain. In the first phase of the interview, the objective was to ensure that all *ST*s were similarly informed (in terms of conceptual comprehension and multi-criteria analysis) before moving on to the second phase. They were invited to produce a 'quality diagnosis' for each explanation by completing the analytical grid (Table 1) for each. Table 2 summarizes the *a priori* quality diagnoses developed by the author for each of the three explanations. For the first six criteria in Table 1, the interviewer was initially non-directive. If the interviewee continued to appear unaware of a flaw in the explanation—for instance, neglect of Archimedes up-thrust and the role of the weight of the immersed object—the interviewer would direct attention to the questionable point. Some examples of this type of discussion can be found in the Results section.

Table 2. The six first criteria and their *a priori* application* for the three explanations.

| | Criteria | Explanation 1 | Explanation 2 | Explanation 3 |
|---|---|---|---|---|
| 1 | Accuracy of the text's conclusion? | Yes (for the contribution of the capillary forces) C+ | Yes (for the contribution of the capillary forces) C+ | Yes (for the contribution of the capillary forces) C+ |
| 2 | Internal contradiction? | No C+ | No C+ | No C+ |
| 3 | Direct contradiction of a law? | No C+ | No C+ | No C+ |
| 4 | Indirect contradiction of a law? | No C+ | Yes. There is no force localised at the triple contact point and tangent to the liquid/gaz interface. C- | No. The capillary forces are shown as distributed over the solid/liquid interface. C+ |
| 5 | Logical completeness? | No. (Archimedes' up-thrust and weight not mentioned.) C- | No. (Archimedes' up-thrust and weight not mentioned.) C- ⟶ The value, direction and localisation of the capillary forces are not justified. C- | No. (Archimedes' up-thrust and weight not mentioned.) C- |
| 6 | Generalisable? | No. (Role of Archimedes' up-thrust?) C- | No. (Role of Archimedes' up-thrust?) C- | No. (Role of Archimedes' up-thrust?) C- |

\* C+ , C-: criterion contributing *a priori* positively (respectively: negatively ) to selection of an explanation.

With regard to the last two criteria (simplicity and mnemonic value), the responses of the *ST*s were briefly discussed, emphasizing that the issue of interest was not an objective feature but the *ST*s' feelings, which were likely to depend on personal experience and conceptual mastery. More generally, the interviewer stressed that the aim of this phase was not to find a binary (yes or no) 'right answer' for each item on the grid but to promote discussion (*de facto*, agreement with Table 2 on the first six criteria was reached with all *ST*s within a relatively short period of time, ≈30').

Directly addressing *RQ1* and *RQ2*, the second phase sought to determine which criteria were decisive for teachers' choices in two contexts: for themselves and for university students, and to what extent they thought such a multi-criteria analysis might be useful in their teaching. During this second phase, the interviewer's role was confined to seeking the *ST*'s opinion in a very neutral way. *ST*s had to choose one of the three proposed explanations, for themselves on the one hand, and for university students on the other hand. In each case, they were asked to justify their decision through the quality diagnoses built in a consensual way during the first phase. They were asked to specify which selection criteria contributed positively (C+) or negatively (C-) to their choices. For this, they were given a sheet as in Table 3 ('criteria sheet': CS).

Table 3. Towards a 'contextualised quality diagnoses': the Criteria Sheet used to collect the interviewees' selection criteria for the three explanations in two contexts *.

| Ad Explanation→ | *Expl. 1* | | *Expl. 2* | | *Expl. 3* | |
|---|---|---|---|---|---|---|
| An explanation: for whom? → | The interviewee | University students | The interviewee | University students | The interviewee | University students |
| ----------------- Selection criterion ↓ | ○ | | | | ⬤ | |
| Criterion in favour of the explanation (C+) | 1, 2, 3, **4** | 1, 2, 3, **4** | 1, 2, 3, **7, 8** | 1, 2, 3, 5, **7, 8** | 1, 2, 3, **4** | 1, 2, 3, **4** |
| Criterion against the explanation (C-) | 5, 6, **7, 8** | 5, 6, **7, 8** | **4**, 5, 6 | **4**, 6 | 5, 6, **7, 8** | 5, 6, **7, 8** |

*In this example, the grid has been completed by a *ST* (Ad); numbers refer to the criteria in Table 1 and a circle designates the explanation that was finally chosen (white for university students and grey for the participant). Items in bold indicate the criteria that the participant said were most decisive.

Interviewees were then asked to state and justify their final choice of one explanation for themselves and for university students. For each criterion and each context, they were asked to underline (in bold, Table 3) which of these judgments were most decisive for their choice. The interviewer again stressed that the survey was about their opinions as young professionals rather than correct answers. Finally, *ST*s were invited to say whether and why they might use this analytic approach to evaluate and select explanations in their teaching practice.

*Processing the interviews*

The first part of the interview was essentially a preparatory phase before data collection. The observations reported below illustrate both the usual limitations of teacher critical awareness and the extent to which the first phase prepared *ST*s to assess the relative merits of the presented explanations, without being affected by a sense of incompetence or poor critical analysis.

In the second phase, interviewees completed their criteria sheets as shown in Table 3. These raw data are grouped in Tables 4 and 5. In addition to these data, relevant comments justifying how they completed the grid are extensively quoted. To address *RQ2*, all *ST*s comments regarding their appreciation of the decision process are grouped by theme and extensively cited. All comments were analysed before their translation in English language.

## 5 Results

*First phase*

Before looking at the implications of the study results for the research questions, it is useful to consider the extent to which the research design created the intended context for the *ST*s at the end of the first phase. This also affords an opportunity to reflect on the kinds of interaction that occurred between the *ST*s and the interviewer.

First, at least 10 minutes elapsed before any interviewee noted that nothing had been said (in *Expl. 1*) about Archimedes up-thrust or the weight of the object. Typical exchanges included the following.

Int $_{27}$ -Nothing else missing? (*Expl. 1, after 4'*)
Ma  -No.
(*7' later*)
Int -Complete? (*Expl. 2*)
Ma -Complete. (…) Generalizable, too.
Int -It's strange that we have not discussed the possible existence of other forces.
Ma -The weight!
Int -And?
Ma -Archimedes' up-thrust!

Less surprisingly, in the case of *Expl. 2*, none of the interviewees spontaneously questioned the representation of the force exerted by the liquid on the solid as acting only at the highest contact point tangential to the liquid surface, before the interviewer decided to clarify this point (after at least 20'). This is probably because this representation is commonly found in textbooks (e.g. de Gennes et al., 2005, p. 43). What follows is an intervention by the interviewer, this time in a very directive style:

Int $_{157}$ - Well, you have a molecule here; how is it pulled? (*Expl. 2, after 20'*)
Th -Well, she's undergoing the force of capillarity.
Int -A force exerted by molecules; return to molecules, forget capillarity—it's nothing more than forces between molecules.
Th -Yeah.
Int -And everything that's there (*in the liquid corner*).
Th -Well, it's gonna pull like that.
Int -Yes, and the total force?
Th -Mm
Int -It can't be tangential, okay?
Th -No, not with these interactions!

Note that when the interviewer stressed the impossibility of a local force 'pulling' on the solid tangentially to the liquid/air interface, the argument was quickly accepted by all. All *ST*s sought further information in relation to *Expl. 3*, and this was provided by the interviewer (after Marchand et al., 2011, p. 1006). All agreed that the important point was that a clear and unequivocal justification was available for *Expl. 3*, based on the distinction between a local interaction near the liquid's edge and a force related to the curvature of the bottom of the cylinder. All were able to distance themselves from the fact that they were not completely comfortable with the mathematical details of *Expl. 3*:

Ju $_{42}$ (*Expl. 3)* – I will have to return on this calculation but I see the main point and the type of argument.

Regarding simplicity and mnemonic value, *ST*s offered diverse reasons for characterizing an explanation as 'simple' or 'easy to memorize'. Concerning simplicity, habits or past experience are mentioned by 5 *ST*s. In the case of *Expl. 1*, all participants emphasised the simplicity of the calculation, but the principle of virtual work was perceived as non-obvious by 4 *ST*s, both for university students and themselves, with remarks such as 'It's misleadingly simple' (Jb $_{84}$) or 'It hides a lot of physics.' (Ju $_{72}$).

To sum up, the six first criteria in Table 2 exposed the usual limitations in *ST*s' critical analysis (i.e. undetected incomplete explanations, an inhibiting effect of habits, exclusive attention to mathematical precision). However, their interaction with the interviewer (framed by the criteria grid) prompted sufficient critical awareness and conceptual mastery to ready them for the second phase. After half an hour at most, the goal of Phase 1 was achieved—that is, when asked whether they had enough information and were sufficiently aligned with Table 2 to discuss how they would choose an explanation, all agreed and no-one anticipated that they would be blocked by feelings of incompetence during Phase 2.

*Phase 2: Selected criteria*

When *ST*s were invited to explain how the criteria positively or negatively influenced their choice of explanation, all began by distributing the criteria in the boxes within each column of the Criteria Sheet. During this process, no participant reconsidered or changed the decision made previously when completing the grid in Table 2, with the exception of criterion 5 (completeness/incompleteness) as detailed below. Having done that, all observed that criteria 1, 2, 3 were positive (C+) and that criterion 6 (generalisability) was negative (C-) for their choices in all columns—that is, for the three explanations, for themselves or their students. They concluded that all of these criteria were non-discriminatory in the present case. For that reason, *RQ1* is addressed in terms of four criteria (4, 5, 7, and 8, see Table 2): presence/absence of an indirect contradiction of a law or of logical incompleteness, simplicity and mnemonic value. Tables 4 and 5 synthesise interviewees' responses in this regard. As three advocated non-dichotomous coding, special symbols ('+' or '**p**') were used to indicate (non-decisive or decisive) judgments that an explanation was considered incomplete but 'less incomplete than *Expl. 2*'.

Table 4 Criteria deemed relevant or decisive in the participant's choice of explanation for themselves

| Explanation | Indirect contradiction of a law? | Logical completeness? | Simplicity? | Mnemonic value? |
|---|---|---|---|---|
| *Expl. 1* chosen by : Ma, Ad, Ju, Au | **p p p p p p** + | - - **n** - - - '+' | **p n** + **p** 0 **p n** | - - **p p** - - + |
| *Expl. 2* | **n n** - - **n n n** | **n** - **n** - - **n** - | **p p** - **p** + **p** + | + + + **p** 0 + + |
| *Expl. 3* chosen by : Th, Pi, Jb, | **p p p** + **p p p** | **n** -'**p**''**p**'- **n**'+' | - - 0 + - **n** + | - - - - 0 - - |

+, **p**: criteria in favour of an explanation, relevant (+) or among the most decisive (**p**)
-, **n**: criteria against an explanation, relevant (-) or among the most decisive (**n**)
'+' '**p**': less negative than for *Expl. 2*
0: irrelevant
In all boxes in columns 2 to 5, symbols referring to each *ST* are in the same order
(1: Ma; 2: Ad; 3: Th; 4: Ju; 5: Au; 6: Pi; 7: Jb).

Table 5 Criteria deemed relevant or decisive in the participant's choice of an explanation for university students

| Explanation | Indirect contradiction of a law? | Logical completeness? | Simplicity? | Mnemonic value? |
|---|---|---|---|---|
| *Expl. 1* chosen by : Au , Pi , Jb | + **p p** + **p p p** | **n** - **n** - - - '+'' | **n n p n** + **n p** | - - **p** - - - **p** |
| *Expl. 2* chosen by : Ma , Ad , Ju | **n n** - - **n n n** | **n** + **n** - - **n** - | **p p p p p p** + | **p** + **p p** + **p** + |
| *Expl. 3* chosen by : Th | **p p p** + **p p** + | **n** -'**p**' '**p**'- -'+' | **n n n n** - **n n** | **n** - **n n** - - **n** |

Notations: see Table 4.

For *RQ1*, the following are the main findings from Tables 4 and 5. Table 4 indicates that participants unanimously valued for themselves what might be called the 'consistency' of *Expl. 1* and *Expl. 3*—that is, the absence of indirect contradiction of a law—along with their relative completeness, despite the simplicity and mnemonic value of *Expl. 2*. In relation to university students, however, Table 5 indicates that the last two criteria – simplicity and mnemonic value - proved decisive for three participants who favoured *Expl. 2*, even though this explanation was deemed inconsistent and broadly incomplete. As a first response to *RQ1*, then, it seems that recognition of an explanation's inconsistency is not sufficient to banish it from *ST*s' pedagogical toolkit.

Beyond these quantitative indicators, *ST*s' comments further illuminate the reasons for their decisions.

*Phase 2: Participants' comments*

A first striking fact is *ST*s' interest in the consistency (and sometimes the 'beauty') of their preferred explanation(s).

Ju $_{204}$ -(*Expl. 3*) Yes for me...I am very happy to have discovered the last one; I am happy to see that there is a less simplistic explanation.
Pi $_{116}$ -(*Expl. 2*) (*laughs*) Yes, well I'm quite happy to see it. (*laughs*)
Pi $_{156}$ -(*Expl. 2*) I like it—it's fine.
Ju $_{174}$ -(*Expl. 1*) Yes I see, it's just a difference in energy; it's nice (…) finally there I see it clearly—it's very simple.
Th $_{242}$-Finally, *Expl. 3* is more complete—complex too. But something is simpler if you understand.
Au $_{170}$-I think criterion 4 (*presence/absence of an implicit contradiction of a law*) is mandatory (*Int - And for undergraduate students now?* ) - It is also criterion 4.
Ma $_{122}$-What seems most important to me is that this one (*Expl. 3*) is not contradictory. I think this criterion is important.

Despite the strong appeal of *Expl. 2*'s simplicity, only three participants chose it for university students. This should not distract from the complexity of participants' internal debate, as evidenced by their comments about completeness and simplicity.

As might be expected, the criterion of completeness was deemed (more or less) negative *a priori* for the three explanations. However, when choosing an explanation for university students, a certain degree of incompleteness was sometimes considered preferable.

Int $_{221}$- So/ incomplete/ It was not explained why the free body diagram was like this.
Ad - That's the downside, actually we're being given it. (…) The minus side is the minus side for me but the plus side for teaching. As to students, we won't necessarily explain everything to them to show how it works. We can omit small details that students won't notice. (…) For teaching, it may be an advantage to omit/
Int - So (*criterion*) 5 is "incomplete", huh?
Ad - Yes, but precisely, in teaching, I think it's good not to go too far or to provide recipes.
Int - So it's incomplete, but it's good?
Ad -That's it.

Pi $_{172}$ - (*Expl. 3*) For me (*completeness*) it's important, for them maybe not so much. Too many questions arise—it's not simple enough.

That said, this type of comment does not necessarily indicate a firm position regarding completeness, as the following paired comments show (same participant for each pair).

Au $_{130}$ - (*Expl. 3*) No, I would have identified incompleteness as a good point because it doesn't interfere with the explanation, (…) fewer parameters are taken into account.
Au $_{150}$ -If it's for an exercise, it doesn't bother me that we forget things; but if it's to understand then (…) no, it doesn't work (...)

Ju $_{220}$ -In fact it still bothers me/ the incompleteness for the students—the weight and Archimedes—yes, because it's one of the first things, they never forget, and then all of a sudden it disappears.
Ju $_{250}$ - For the students, it doesn't bother me if we go against a physical law (…) we can't say everything.

With regard to simplicity, three participants said this criterion was not decisive for themselves.

Ju $_{194}$ (*Expl. 1*) - For myself, it doesn't matter that it's not simple.

Th $_{242}$ - Ultimately, *Expl. 3* is more complete and complex, and I find it a little less simple, but this is not against (*it*) because something is simpler if you understand.

In selecting an explanation for students, some participants were ready to compromise temporarily between consistency and simplicity:

Th $_{262}$ - (*criterion 4*) It's obviously important from a scientific point of view, but it depends on how I see it; for students who are in difficulty, for example, I prefer to teach them first that the force goes in that direction *(as in Expl. 2)* and then come back to it once they've got it.

This aligns with a final claim that simplicity should not be seen as a paralysing necessity.

(*Int: On the side of simplicity, do you think there are still some things that are annoying about Expl. 3?*)
Au $_{180}$ - Yes, but they are able to understand. (*You are telling me simplicity first, but then…*) - It's not that complicated.
Th $_{296}$ - Well, I don't see why they couldn't understand what I understood.

A remarkable change of opinion during the interview shows how difficult it is for participants to reconcile consistency, completeness, and simplicity.

Pi $_{177}$ - (…) We spend our time giving students incomplete explanations, so uh, no, for me it's not a problem (*laughing*) that it's incomplete.
Pi $_{196}$ - (*Expl. 2*) For students, I'm going to take this one, and I hate myself because I've always hated that explanation. (…). In the end, I'm not going to hate myself, and I'll take *Expl. 1* for the students. After discussing, yes go, that one. (*The one for students?*) That's right, number 1.

In summary, beyond the quantitative elements in Tables 4 and 5, these comments confirm both *ST*s' interest in consistency and their hesitations as to the place to be given to simplicity and completeness when choosing an explanation.

*Phase 2: Benefits of the proposed approach*

When asked to assess the proposed multi-criteria diagnosis approach to choosing an explanation, *ST*s were unanimously very positive, with some nuances. All valued how this approach facilitated deliberate and well-informed decisions.

Ju $_{248}$ - Right now, we have criteria that enable us to reflect, and after making choices … It enables me to clarify my way of thinking and my preferences (…) I know what I do and why I do it, whereas I didn't know before. The analytical approach, it helps a lot, yes it helps a lot.

Ad $_{282}$ -Clearly, for me, it is a very good method if not the best possible (…). It makes me think about what I want to do when teaching. It's really not bad! No, I've never done it to that extent, but now I see all the advantages. I must admit I don't really see a downside.

Some metacognitive comments referred to the nature of science, stressing the fact that 'there is no perfect explanation' or the importance of consistency:

Pi $_{214}$ - As I already know, what strikes me is that there is no perfect explanation when I still have the story in my head that physics is simple.

Pi $_{228}$- For that reason, this analysis by criteria is interesting. I learned that I assign a lot of importance to this criterion (*consistency*). I already knew it a little, but this confirmed it. It's very interesting.

Others stressed that while this approach provides useful information, it is not a decision-making algorithm.

Au $_{186}$ - It allows you to see what's important, so it's good … It eliminates some explanations, It's true—when you have to choose, you can't do everything, and then I couldn't decide between two explanations.

Pi $_{218}$ - In fact, it helped me to complete this analysis step by step in order to see the defects, the advantages of each—for sure, it helped me to do the analysis, but I didn't base my choice mathematically on these criteria.

Finally, and strikingly, two STs who expressed difficulty in choosing one explanation gave very precise reasons for each possible choice.

Au $_{164}$ - For me, I would have taken *Expl. 3* and then *Expl. 1* (…) I would take *Expl. 3* to understand well, but also *Expl. 1* to understand the formalism of energy and to do exercises.

Ju $_{210}$ - I'm fine with all three, actually, for different reasons. (…) Yes, I like the first one because it is beautiful; the second because it is intuitive; and the third—I like it because it seems more correct. They all have a real strength.

It was also suggested that this analytic approach might facilitate and enhance discussions between colleagues by introducing more detailed and explicit arguments.

Ma $_{158}$ – Between colleagues it has the advantage once again of putting things in perspective and that what we have in our heads is visible to everyone. In the context of a discussion, I think it is always important to communicate well with the other person.

# 6 Recapitulation and discussion

This investigation addressed two research questions concerning *ST*s' decision-making process for the choice of their explanations in physics: When several explanations of a given physical phenomenon are available and have been critically analysed based on a list of criteria, which of these criteria are decisive in *ST*s' decision making when they choose an explanation in their teaching?; after making a choice based on multi-criteria analysis of several explanations, how did *ST*s assess the value of this approach for their teaching practice?

Seven *ST*s were interviewed individually and were presented with three explanations and a list of assessment criteria. The first phase was devoted to establishing a consensual multi-criteria evaluation or 'quality diagnosis' of each explanation to ensure that *ST*s moved on to *Phase 2* with a broadly similar conceptual understanding and critical analysis of the three explanations. During the second phase, *ST*s were asked to decide which criteria were decisive for their choice of explanation (for themselves and for university students), to choose one explanation for each target and to comment on the value of the multi-criteria decision-making process.

The goal of *Phase 1* was completed as planned; after half an hour, all interviewees were in possession of more or less the same assets (i.e. comprehension and critical analysis) for deciding which explanation to use in a given context. The second interview phase centred first on *RQ1*, regarding *ST*s' decision criteria following thorough critical analysis of the candidate explanations—for instance, would they exclude an explanation on realising that it was not consistent with a physical law? The interviewees were invited to return to the diagnostic grid to choose one of the three explanations for themselves and for university students. During this process, their judgments aligned with those on the *a priori* grids, with the exception of 'incompleteness', which was sometimes identified as a positive factor.

Consequently, the final discussion related to 'consistency' (compatibility with accepted laws), as well as the completeness, simplicity and mnemonic value of each explanation. In the discussions preceding their final choices, two issues emerged. Firstly, the responses of the *ST*s suggest that, for themselves, they unanimously appreciate the consistency (*Expl.1* and *3*) and relative completeness (*Expl. 3*) despite the advantages of simplicity and mnemonic value of *Expl. 2*. In contrast, three participants referred to these two criteria in selecting *Expl. 2* for university students, even though this explanation was deemed inconsistent and broadly incomplete. In other words, an explanation deemed incompatible with accepted laws was not necessarily banished from teaching practice. Incompleteness was deemed even less relevant as an exclusion criterion; indeed, several *ST*s considered it necessary in the case of university students. Given the thorough preceding critical analysis, these results cannot be attributed to a lack of awareness of the participating *ST*s. In relation to current teaching practice, this means that not all decisions to use an inconsistent or very incomplete explanation are due to defective critical analysis, even if this limitation may contribute to their choices.

On the other hand, it may seem surprising that four of the seven *ST*s chose *Expl. 1* or *Expl. 3* for university students despite the complexity of those explanations. This confirms the great importance ascribed to consistency. However, it would be unwise to attempt any more general quantitative prediction of *ST*s' decisions about these three explanations, given the very small sample and the fact that interaction with an interviewer may have contributed to the strong concerns about consistency. That said, *ST*s' metacognitive comments reveal some views which are probably less context-dependent. In particular, their comments revealed conflicting and unstable positions regarding the value of consistency and simplicity in choosing an explanation. Finally, commenting on the approach proposed during these interviews, all responded very positively. They deeply appreciated the multi-criteria evaluation of proposed explanations, even if this did not amount to a decision-making algorithm. In particular, they saw great value in knowing what they are doing, and why, when deciding on an appropriate explanation for a given audience. It was also suggested that more explicit arguments might facilitate discussions with colleagues.

## 7 Final remarks

Although this was not explicitly stated, it seems possible that *ST*s' satisfaction with the proposed approach was linked in part to a sense of appropriating their own teaching decisions. As observed in previous studies, both habits and feelings of incompetence contribute to critical passivity, and expert anaesthesia contributes to not seeing an explanation's flaws by unconsciously correcting its contestable aspects. The fact that implicit rules (such as the imperative of simplicity) are rarely discussed may also limit reflection. In contrast, the type of interactive activity reported here invites teachers to reconsider and analyse both well-known and less familiar explanations. Realising that explanations of a given phenomenon are not univocal or undisputable, even for domain experts, could lead *ST*s to a more active role of responsible decision-maker. Even where national directives or other contextual factors constrain explicative choices, this more thoughtful approach is likely to enrich *ST*s' practice. At the very least, such activities promise to enhance awareness, which is in itself a valuable outcome.

That said, there is a need to better understand how to facilitate teachers' multi-criteria evaluation of alternative explanations. In particular, the idea of 'simplicity' is not straightforward and warrants closer attention in the context of science education research and teacher preparation. Is simplicity the primary requirement for students, and is this purely a matter of reduced mathematical complexity? Regarding incompleteness, does avoiding mention of all underpinning hypotheses make an explanation easier to understand?

In particular, it would be worth investigating the consequences of the widespread assertion that modelling is first and foremost a matter of simplification. This may unduly legitimise the emphasis on simplicity in model construction, implying that models by their very status are somehow immune to criticism. The interview-based approach described here seems to have engendered a salutary awareness in this regard: 'I have the impression that, when we try to simplify, we completely lose the meaning of what we try to explain.' This *ST*'s comment deserves to be taken seriously, as it underlines the difficulty of finding the right balance between simplicity and consistency for a given audience. In terms of teacher preparation, this comment reflects a personal learning style that Vermunt (1996) calls 'meaning-oriented'. As this issue arose following a long critical discussion of contestable explanations with an interviewer, it seems reasonable to infer that this activity played some part in the insight. Similar metacognitive comments in the present study—'I learned that I assign a lot of importance to this criterion (consistency)' or 'It's a misleading simplicity'—suggest that this type of critical discussion is worthwhile for teacher preparation. In general, this highlights a distinction between simplifications that do not compromise the essence of explanation and others that are somehow 'toxic'. For example, accounts like the notorious 'isobaric hot air balloon' ruin the very foundation of appropriate explanation.

On this basis, it is reasonable to suggest that teacher preparation courses should encourage critical analysis of explanations based on multiple criteria. Students may be invited to criticize their own analysis of a given phenomenon. However, this does not extend to many of the explanations currently used in teaching, and it is rare to discuss issues such as the value of simplicity and consistency with beginning teachers. This is probably because there is no simple decision-making algorithm for explanatory choices, which is a further reason for providing *ST*s with practical tools and encouraging reflection to equip them as well-informed, thoughtful, and responsible decision-makers. In the very limited context of this study, the potential benefits of activities that link critique and decision-making seem significant.

That said, teachers' access to relevant critical analysis of physics explanations is an important prerequisite for such activities. This was positively facilitated by the interviewer, the *ST*s being autonomous only for the selection of criteria that were decisive in determining pedagogical decisions. It is reasonable to infer that the teachers would have struggled to explicate their key criteria in the absence of this assistance. In current practice, teachers are *de facto* in charge of both of these intellectual activities. However, despite the multitude of available studies of critical thinking, little is currently known about how teachers' critical capacity can be encouraged more effectively. In this regard, it seems likely that fruitful preparation depends in part on preliminary conditions that help teachers to overcome psycho-cognitive blockages such as feeling of incompetence or addiction to teaching rituals. To that end, it seems essential to explicate the critical dimension in teacher preparation by emphasising that science teaching should focus on comprehension and its twin 'legs' of understanding *and* critique. Useful activities might include co-construction of thorough content analyses with an expert or other *ST*s, preferably using an analytical grid. At the same time, it is important to note that efficiency in this domain is likely to require a long-term approach. It also seems likely that inviting *ST*s to discuss and explicate their decision-making criteria (as in the present study) can serve to motivate the development of critical ability.

Further research is clearly needed to more precisely document the difficulties and positive outcomes of this form of interaction in teacher education. Among theoretical perspectives that might inform such a research program, TSPCK is a priori a good candidate for further development, despite the difficulties noted by Mavhunga and Rollnick (2013). In this regard, Braaten and Windschitl's (2011) advocacy of 'a conversation (…) in the field about how particular scaffolding supports for scientific explanations are useful for supporting the development of students' explanatory ideas *as well as* argumentation practices' seems apt; developing 'conversations' across the lines of research on argumentation, explanation, critical thinking and PCK (e.g. Tang et al. 2020) seems a useful strategy for further exploration of how teachers make decisions about which explanations they will use.

Chabay, R. & Sherwood, B., (2006), Restructuring the introductory electricity and magnetism course, *Am. J. Phys.* 74, 329–36.

Chauvet, F. (1996). Teaching colour: design and evaluation of a sequence *Eur. J. Teach. Educ*. 19, 119–34.

Chevallard, Y. (1991). *La transposition didactique*. Grenoble: La pensée sauvage

de Gennes, P. G., Brochard Wyart, F. & Quéré, D. (2005). *Gouttes, bulles, perles et ondes* : [*Drops, bubbles, pearls and waves*], Paris: Belin.

Décamp, N. & Viennot, L. (2015). Co-development of conceptual understanding and critical attitude, Analysing texts on radio-carbon dating, *International Journal of Science Education*, 37 (12), 2038-2063. http://dx.doi.org/10.1080/09500693.2015.1061720.

Duschl, R.A. & Osborne, J. (2002). Supporting and Promoting Argumentation Discourse in S*cience Education*, , 38:1, 39-72, DOI: 10.1080/03057260208560187

Ennis, R. H. (1992). The degree to which critical thinking is subject specific: Clarification and needed research. In S. Norris (Ed.), *The generalizability of critical thinking: Multiple perspective on an educational ideal* (pp. 21–37). New York, NY: Teachers College Press.

Etkina, E. & Planinšič, G. (2015). Defining and Developing 'Critical Thinking' Through Devising and Testing Multiple Explanations of the Same Phenomenon, *The Physics Teacher* 53, 432-437. doi: 10.1119/1.4931014

European Commission (2015). *Science education for responsible citizenship,* Report EUR 26893 (EN chair H. Hazelkorn, Brussels).

Gess-Newsome, J. (2015). A model of teacher professional knowledge and skill including PCK: Results of the thinking from the PCK Summit. In A. Berry, P. Friedrichsen, & J. Loughran (Eds.), Reexamining pedagogical content knowledge in science education (pp. 28–42). London: Routledge.

Härtel, H. (1985). The electric circuit as a system, Aspects of Understanding Electricity, In R. Duit et al. (ed), Keil: Schmidt & Klaunig, pp 343–52.

Hawthorne, R. K. (1986). The professional teacher's dilemma: Balancing autonomy and obligation. *Educational Leadership*, 44(2), 34-35.

Hecht E. (1996). Physics Calculus Brooks/Cole, International Thomson Publishing.

Hempel, C. G. (1965). Aspects of scientific explanation. In C. G. Hempel (Ed.), Aspects of scientific explanation and other essays in the philosophy of science. New York: Free Press.
27